\documentclass{article}

\usepackage{multirow}
\usepackage{graphicx}

\begin{document}

\title{Gender of Recruiter Makes a Difference: A study into
 Cybersecurity Graduate Recruitment}

\author{Joanne L. Hall and Asha Rao\\
Department of Mathematical Sciences, School of Science, \\RMIT University, Australia}

\maketitle

\begin{abstract}
An ever-widening workforce gap exists in the global cybersecurity industry but diverse talent is underutilized. The global cybersecurity workforce is only 25\% female. Much research exists on the effect of gender bias on the hiring of women into the technical workforce, but little on how the gender of the recruiter (gender difference) affects recruitment decisions. This research reveals differences between the non-technical skills sought by female vs non-female  cybersecurity recruiters. The former look for recruits with people-focused skills while the latter look for task-focused skills, highlighting the need for gender diversity in recruitment panels.

Recruiters are increasingly seeking non-technical (soft) skills in technical graduate recruits. This requires STEM curriculum in Universities to adapt to match. Designing an industry-ready cybersecurity curriculum requires knowledge of these non-technical skills. An online survey of  cybersecurity professionals was used to determine the most sought after non-technical skills in the field. Analysis of the data reveals distinct gender differences in the non-technical skills most valued in a recruit, based on the gender of the recruiter (not the recruited). The gender differences discovered do not correspond to the higher proportion of women employed in non-technical cybersecurity roles.
\end{abstract}


\section{Introduction}
Could the gender of the recruiter affect recruitment decisions? Plenty of evidence exists of bias in STEM recruitment relating to the gender of the person being hired \cite{ moss2012science,Gorsuch2019gender}.  Anonymization of job applications can reduce this type of gender bias in recruitment   \cite{HBRanonymization2020}.
Little research, however, exists on the effect the gender of the recruiter (\textit{bias from gender or gender difference}) may have on recruitment decisions. This makes it hard to qualify or quantify the contribution of gendered panels to gender bias, or, alternatively, the benefits of gender diversity on a recruitment panel. This study addresses this research gap. 

An anonymous survey of cybersecurity recruiters was conducted \cite{HallRao2020} to ascertain the skills sought by cybersecurity industry professionals  in their graduate recruits. 
Gender demographics collected as part of the survey, as suggested by \cite{UN2016}, resulted in these findings, which indicate that the gender of the recruiter could  impact  the decisions made with regards to the skills sought in graduate recruits. Female recruiters look for people-focused interactive skills in new recruits, while non-females seek recruits with task-focused skills. The number of  respondents  (78) to this survey  is small, but the $p$ value of 0.0034 indicates that the effect found in our data is extremely unlikely to be due to chance. Gender differences could be influencing recruitment in the cybersecurity space.

The socio-technical nature of cybersecurity  requires its practitioners to not only have technical and domain specific knowledge but also social intelligence to be successful \cite{dawson2018future}. 
Despite the prevailing technical image of cybersecurity,  most successful cyber attacks involve  human-centric methodologies including phishing and misuse of privileges \cite{verizon2022}.   Non-technical skills such as teamwork and communication are essential for cybersecurity professionals, with many graduates needing intense on-the-job training before starting work \cite{crumpler2019cybersecurity}.
Non-technical (soft) skills are often listed along with technical skills in many job advertisements in cybersecurity. Research shows that professionals in the field consider these skills necessary \cite{HallRao2020, GroenVenAerts2021,Suss2021}. 

The survey \cite{HallRao2020} capturing the data analysed in this paper, was designed to specifically determine the needs of the  cybersecurity industry, with respect to the non-technical skills sought in graduate recruits. The demographic data of all survey participants was collected to ensure a suitably representative sample \cite{UN2016} of the cybersecurity industry. Preliminary analysis \cite{HallRao2020} identified curiosity and creativity as the non-technical skills most valued by cybersecurity practitioners; most Australian cybersecurity degree curricula include neither explicitly. 

In this paper, the survey data is analysed using gender as a primary identifier  and the following research question is examined:

\textbf{Research Question:} \emph{Are there significant gender differences in the non-technical skills sought when recruiting cybersecurity graduates?}\label{RQ}

\section{The Cybersecurity Workforce: A Critical Analysis  \label{sec:current}}

We examine the need for this research  with regards to the cybersecurity industry: the workplace culture, current gender balance statistics, the skills shortage, existing classifications of work roles and the cybersecurity curriculum. 

\subsection{Cybersecurity Workplace Culture \label{sec:cul}}
Cybersecurity is a male-dominated technical field \cite{ISC2022}. While workplace cultures differ widely across country, industry, and company, the theme of the male dominated workplace culture keeps recurring \cite{schmitt2021women}.  To succeed in such workplaces women continue to adopt ways of working unnatural to them \cite{Bix2019}.

Often, confidence is valued, but not associated with career progression in women \cite{risse2020leaning}. Professional identity is wound up in the masculine ideal of displaying technical competence, even for roles that do not require any technical tasks \cite{Faulkner2009a}. %

Women are supposedly better at communication and organisational skills, but men more often exploit the informal career progression pathways of networking and bragging about achievements \cite{von2000people}, with overt displays of unrelated technical competence \cite{Faulkner2009a}. Thus, career progression and job opportunities can come from workplace cultural activities rather than the skills and knowledge required for a particular role. 

A diverse, gender balanced workforce and environment has many advantages \cite{DobbinKalev2022}. Some research focusing on the low numbers of women working or studying in cybersecurity, and the barriers they face \cite{shumba2013cybersecurity}  comes from a deficit model \cite{fox2017},  suggesting women need extra support to cope in the male dominated environment.      Neither female dominated nor male dominated behaviour should be considered problematic.  Rather, better outcomes will be achieved by working with teams that incorporate a range of behaviours \cite{poster2018cybersecurity}.

\subsection{Gender in the  Cybersecurity Industry}

The Global cybersecurity workforce is estimated to be about 25\% female \cite{ISC2022}. The Australian one is currently only 17\% female \cite{Risse2023}.  The Australian government, understanding the need to strengthen the workforce, has examined the industry  through the lens of gender \cite{foley2017women}, and suggested actions to recruit and retain women   \cite{AustCyber2018}.   

Mentoring and providing support to women at all jobs levels is acknowledged as necessary 
\cite{ISC2020}. In Australia, several programs support women in technical fields including the AWSN Explorers \cite{AWSN-webpage}.  However, the impact of these actions and projects is yet to be evaluated \cite{mckinnon2020absence}. 

A significant increase in female students undertaking technical education and training could close the cybersecurity workforce gap and improve the gender balance  \cite{AustCyber2018, ISC2020}. The percentage of female students enrolling in technical degrees, including computer science and engineering (common pathways into cybersecurity), is slowly trending upwards in many countries, with undergraduate enrolments now between 15\% and 25\% female  \cite{NCWIT2020}.    

\subsection{Cybersecurity Industry Skills shortage}
 The ever-increasing proportion of global economies  reliant on digital technology  feeds the increased need for skilled cybersecurity staff. However, recruiting suitably skilled personnel can be a challenge. The cross-jurisdictional nature of cybercrime further exacerbates the challenge. 

A high degree of uncertainty in the available data on the composition of the cybersecurity workforce exists due to the unregistered, unlicensed nature of the cybersecurity industry.  This data paucity is starting to be addressed With the  inclusion of the cybersecurity  profession in the census for the first time \cite{ABS2021}. 

The global cybersecurity workforce (around 4.7m) grew by 11.1\% from 2021, with the Australian workforce growing by almost 7\% \cite{ISC2022}. In Australia, women comprise around 17\% \cite{Risse2023} of the cybersecurity occupations (a new category in \cite{ABS2021}). However, the workforce gap grew by over 57\% in Australia \cite{ISC2022} and the percentage of women in Australian ICT  Security Specialisations (the category that includes cybersecurity occupations) fell from 19\% in 2006 to 16\% in 2021 \cite{Risse2023}.
Organisations seeking staff in the highly competitive labour market need to develop better recruitment practices.   

As an unlicenced profession, many pathways exist into a cybersecurity career; almost 40\% of job advertisements don't require university study \cite{marquardson2020skills}. Traditionally, many cybersecurity professionals entered the industry through interest, self-study and on-the-job training. Not-for-profit organisations and training businesses offer training courses and certifications \cite{ISC2020} to such aspirants, but there is no requirement to complete a certification before beginning work in a cybersecurity role. 
Cybersecurity certification is a decreasing trend  \cite{ISC2022} with employers no longer treating certifications as an entry requirement. 
As  the skills shortage deepens,  formal education and training has become a common pathway, with  23\% of under 30s now entering the field after pursuing formal education \cite{ISC2022}.

\subsection{Industry Ready Cybersecurity Curricula}

Prospective students' choice of a master's degree often hinges on the proportion (perceived or otherwise) of new graduates finding relevant work opportunities. Student employment outcomes also influence government accreditation of educational institutions and courses of study \cite{QILT}, and are often used when compiling rankings of educational institutions \cite{QSemploy}.

In 2019,  13 Australian universities offered a master's degree in cybersecurity 
with several other universities offering a cybersecurity major within other master's degrees \cite{HallRao2020}. 
However,  there was little consensus on the skills that should be taught in a cybersecurity course of study, or the skills needed to begin a cybersecurity career \cite{HallRao2020}. Cybersecurity practitioners need non-technical skills such as teamwork, communication and curiosity. However, many cybersecurity programs are mostly technical and neither teach nor assess many non-technical skills.

Accreditation leads to industry advising curriculum. Accredited Engineering degrees require the assessment of non-technical skills such as teamwork and ethical conduct \cite{EA2019}. 
 The Australian Computer Society \cite{webpage-ACS}  recently started offering accreditation for cybersecurity degrees;  a new scheme whose impact is yet to be seen.

\subsection{Non-Technical Skills in a Cybersecurity Career \label{sec:nontechcareer}}

A successful cybersecurity professional must engage with both the technical and non-technical aspects of cyber defence. 
People-focused skills such as teamwork and verbal communication are highly valued in the cybersecurity workforce. The reality is not the popular media image of the lone male in a hoodie working away on a computer in a darkened room. 

Cybersecurity systems satisfy the five characteristics of a socio-technical systems \cite{baxter2011socio}: they have interdependent parts; have separate, but interdependent, technical and social subsystems; must adapt to changes in the external context;  design choices must be made in all parts to achieve a security goal; and their performance relies on the collective and symbiotic performance of both the technical and social subsystems.

A non-technical skill is one that can be demonstrated without the use of technology;   verbal communication can be demonstrated using a video conferencing platform, or two people conversing together at a ca\'{f}e.  Non-technical skills  are also called  `soft skills' \cite{bancino2007soft} or `professional skills' \cite{cerri2016fully}.

Non-technical skills vary; communications skills, for example, differ according to what is being communicated. Communicating technical outcomes within one's own technical discipline is different to communicating with a wider range of audiences over a corresponding range of mediums  -- wider project success, influencing  multiple, diverse teams, and ultimately, career advancement depends on the latter \cite{cerri2016fully}. Someone with strong technical communication skills may lack the ability to communicate with those outside their technical discipline.  Thus, simply listing `communication skills' in a job advertisement may not attract the skill set required by the hiring manager \cite{riemer2007communication}.  

Cybersecurity comes with its own unique challenges. An in-depth understanding of technology is necessary to defend an information system, but project management and teamwork, communication and leadership, creativity and flexibility are just some of the  skills necessary to deliver a successful cyber-defence project. Specialist skills in governance, management and coordination are needed to address systems-level challenges and meet the industry’s large-scale cybersecurity needs  \cite{Risse2023} .

\subsection{Cybersecurity Work Roles}\label{workrole}
The US National Institute of Standards and Technology (NIST)  attempted to address the workforce gap in cybersecurity with the NICE Framework \cite{NICE2017}, providing knowledge, skills and abilities needed for 52 cybersecurity roles within an organisation. 

While the NICE Framework  has support from international governments, it is aimed at large organisations and oriented towards the defence industry \cite{Canada2023}.  It provides scant information on the non-technical skills needed by cybersecurity practitioners \cite{Suss2021}.

This is in contrast to industry reports \cite{Bate18}, surveys \cite{Tripwire2017} and research \cite{HallRao2020,Suss2021}. In Tripwire's survey \cite{Tripwire2017}, 72\% of respondents confirmed the need for soft (non-technical) skills such as analytical thinking and communication as having increased over the previous 2 years. Employers showed willingness to hire people with strong non-technical skills but little technical expertise. 

In 2020  the Workforce Framework for Cybersecurity \cite{NICE2020} replaced \cite{NICE2017}, presenting instead ``building blocks'' of tasks, knowledge and skills. Work roles are no longer listed. 

\subsection{Current Research Gaps}

Graduate recruitment is a possible pathway to reducing the workforce gap. 
Prior research on non-technical skills (see section \ref{sec:nontechcareer})  does not consider graduate recruitment, an important career stage. The Frameworks, including the NICE Framework, do not list any of the work roles as suitable for new graduates.  Understanding graduate recruitment is also important for universities; it  helps shape the experiences of people prior to and during the graduate recruitment phase \cite{QILT}.

The gendered data from technology-based industries \cite{NCWIT2020,Industry2023}  is often in terms of the disproportionately few women in technical roles, or in  leadership.   Some recent research exists on diversity in recruitment using  USA \cite{lyon2020exploring} and European \cite{corneliussen2020brings} data sets, as well as some decades-old research in an Australian context \cite{trauth2003explaining, von2000people}.  These studies explore the entire career pathway, including attraction, recruitment and retention of women and people of minority background.   Diversity in hiring panels is often mentioned as being important in the recruitment of diverse applicants \cite{Fuhl2020-eliminatinggenderbias}, but little research exists about this aspect of recruitment.

Much of existing literature on the University/workplace transition of graduates focuses on the experiences and perspectives of early career professionals \cite{ lutz2021exploring, brunhaver2021early}.
This study seeks to understand the perspectives of those who lead, mentor and manage early career staff in the cybersecurity domain. The differences occurring in the recruitment phase are examined, determined by the gender of the recruiter,  without taking into consideration the gender of those being recruited.  This is a subtly different question and is important in understanding gender diversity in the cybersecurity industry. 

This research uses an Australian data set.  Whilst studies and data collected in culturally similar countries are often applicable to the Australian context \cite{shenkar2001cultural},  Australia's geographic isolation can create aberrations, as is evident from the lower percentages of female cybersecurity professionals in the   Australian vs the global cybersecurity workforce. 

Understanding the non-technical skills sought by industry and the effect of gender differences on these skills, combined with the on-going conversation  on cybersecurity work force roles would help in understanding how to address the cybersecurity workforce gap. 

\section{Methodology \label{sec:method}}
An anonymous online survey was conducted in mid-2019 to identify and understand the non-technical skills sought in graduate recruits  by professionals in the Australian cybersecurity industry. 

The preliminary analysis of the data \cite{HallRao2020} compared the skills sought by survey respondents (employers) with the skills  taught in Australian universities in Master of Cybersecurity degrees. It revealed a  significant mismatch with skills, such as teamwork, being highly sought after in industry yet rarely explicitly taught or assessed in these postgraduate degrees.  

In this paper, further analysis of the same dataset is presented, examining the gender differences in the skills being sought in cybersecurity graduate recruits. We answer the  research question presented here (page \pageref{RQ}) based on the answers to the questions posed in the survey. 

\subsection{Data Collection}
We start with the demographics collected in the survey, and then move on to the questions asked.  
\subsubsection{Survey Demographics}
 A range of demographic data was collected, including gender, location, and role, to understand the breadth of the survey participants and compare with known statistics. Table \ref{tab:demographics} gives the summary statistics from the survey compared with available statistics. 

 \begin{table}[h!]
    \centering
    \caption{Demographic and comparison statistics of (2019) survey. Table gives categories and percentages of survey respondents, as compared to known percentages in literature.}
    \begin{tabular}{lrl}\hline
     Survey Respondents & \% resp. & Comparison statistics (\%) \\ \hline
         Working in Australia & 82 & \\
         Female & 19 & 17 Australia \cite{Risse2023};\\ 
         & & 25 Global \cite{ISC2020};
         \\
         Technical Role & 71 & 
         \\
         Non-technical Role & 29 & \\
         Female Technical& 13 & \\
         Female Non-technical & 24 &\\
         \hline
         
    \end{tabular}
    
    \label{tab:demographics}
\end{table}

 Of the 78  survey respondents reporting as working in cybersecurity related roles,  64 (82\%)  stated that they worked in Australia.  
 
 Participants in the survey were asked to identify themselves as female via a yes/no (binary) question. Respondents were given the option to choose ``prefer not to say".  The term non-female is  used in this research to include anybody not identifying themselves as female, to enable comparison with Australian government statistics, which, in 2019, only contained binary gender identification. Furthermore, more recent Australian and global  research still describes the percentage of non-binary respondents as too small to enable meaningful analysis \cite{ISC2022,Risse2023}.  
With a gender split among the respondents of  81\% non-female and 19\% female (in keeping with stated  nomenclature), the survey population reflects  Australia's current ICT security workforce \cite{Risse2023}.  
  
 Respondents were asked to self-identify as currently having a technical or non-technical role. Of the survey respondents, 29\% self-identified as having a non-technical role.

 There is no research on the split of technical vs non-technical roles within cybersecurity. Neither does NICE framework (2017) break down its listed roles in this manner. Furthermore, there exists no data on the number of people employed in the various work roles listed in the NICE Framework within Australia (or even  what proportion of these roles currently exist in Australia). 
 
Among those identifying themselves as currently or recently employed in technical roles, 13\% identified themselves as female. Of those employed in a non-technical role, 24\% identified themselves as female. 
This is comparable to the trend of women working in technical fields being more likely to be working in non-technical roles \cite{NCWIT2020}.   The female respondent pool was evenly split between technical and non-technical roles, whereas the non-female respondents were two thirds technical and one third non-technical. 

\subsubsection{Survey Questions}
The demand for recruits into technical positions to have non-technical skills has increased with the outsourcing of technical roles \cite{bancino2007soft}. Thus, it was of interest to check the match and mismatch between the skills used by recruiters themselves and those they thought as necessary in recruits. Accordingly, survey participants were asked about both the non-technical skills they used in their roles as well as the non-technical skills they sought when hiring junior staff, primarily graduates without industry experience. Survey participants were given a list of skills (Table \ref{tab:SkillList}) to choose from. They were also given the opportunity to add skills that were not in the list, in a free text box. 

\begin{table}[h!]
    \centering
 \caption{Non-technical skills listed in survey. Respondents were asked to select up to 3 skills from this list.}   
 \begin{tabular}{ll}\hline
    Written Communication & Leadership and social influencing\\
    Verbal Communication & Personal Integrity\\
    Flexibility and Adaptability & Work Ethic \& Personal Organisation\\
    Curiosity and Creativity & Emotional Intelligence\\
    Confidence and Positive Attitude & Inclusivity and Teamwork\\
    \end{tabular}
       \label{tab:SkillList}
\end{table}

The list of skills provided (see Table \ref{tab:SkillList}) to survey participants was, in the main,  adopted from the 2018 Future of Jobs Report \cite{world2018future}. In the  report, the  following non-technical skills were listed as either in high demand (2018), trending to be in high demand by 2022, or being in decline by 2022.

\begin{itemize}
\item  Leadership and Social Influencing (Current and Trending),
\item Verbal Communication (Declining),
\item Personal Integrity (Current),
\item Work Ethic and Personal Organisation (Current and Declining),
\item Curiosity and Creativity (Current and Trending),
\item Emotional Intelligence (Current and Trending).
\end{itemize}

Given that the survey was conducted in mid-2019, we compare the skills listed above with the more recent 2023 Future of Jobs report \cite{world2023future}. The skills have now been placed into various categories. Creativity,  a cognitive skill, and curiosity, a self-efficacy skill, continue to be in the top 5 most important skills for workers. Leadership and social influencing,  a working-with-others skill, remains in the top 10 list.

 An examination of Australian cybersecurity job advertisements  \cite{HallRao2020} indicated the above list was not comprehensive, and the following skills were added to the survey list.
\begin{itemize}
\item Written Communication,
\item Confidence and Positive Attitude,
\item Flexibility and Adaptability,
\item Inclusivity and Teamwork.
\end{itemize}

The skills listed in the survey (Table \ref{tab:SkillList}) were not demarcated in any way to differentiate between those taken from the  \cite{world2018future} report, and those obtained by scanning job listings.
In addition, a free text box was provided to survey participants to add skills that were not on the list, but which they felt were important enough to be considered.

The following two questions were asked of all survey respondents:
\begin{itemize}
\item[]Question A: {\em In your current or most recent professional role,
what are the most important non-technical skills?}
\item[]Question B: {\em When recruiting junior staff into your team, what
non-technical skills are you most looking for?}
\end{itemize}
Each respondent could select up to three skills from the list in Table \ref{tab:SkillList} for each question and write up to 250 words in the free text box associated with each question.  Over half of the survey respondents wrote additional comments in the free text boxes.  These comments were categorised  and included in the survey results.

Questions A and B are similar, but also different. Question A asks survey respondents about their personal use of non-technical skills, and hence, is analogous to  methodologies of other qualitative studies \cite{trauth2003explaining, jesiek2021performing}  on career progression. This question is important as it helps understand if biases in skills are related to the role type of the survey population, that is, if people in  technical roles give importance to different skills than those in non-technical roles.  

Question B, on the other hand, looks for the skills the survey respondents would want in a junior recruit. The survey respondents were assumed to already be  in the  workforce, and hence not in the pool of people related to Question B (possible recruits).

Differences in answers to Question A versus B could indicate differences in work experience, role type or evaluation of self vs employee.  Separate analyses of the answers to the two questions, using demographic data, can indicate whether experience is the more important factor, or the demographics of the survey  population. Plenty of research (eg. \cite{Tsrikas2020perceptions}) studies  the different perceptions of non-technical skills between employees and employers. However, little research exists on whether recruiters look for skills different to their own in potential recruits.

Since the main purpose of the survey  was to design industry-ready higher education curriculum in cybersecurity, the data from the answers to Question B took precedence.

\subsection{Data Analysis} \label{subsec_dataanalysis}

 Both questions A and B were related to skills, without any reference to individual attributes such as the gender, race or age of the person possessing those skills. Neither of the survey questions shed any light on bias towards gender.    Hence any gender bias  comes from the gender of the respondent. That is, gender differences within the \textit{recruiting} cohort are examined.

Bias towards something could  develop from a mixture of past educational, professional, and personal experiences.  Bias can also be induced by controlling the experiences just prior to, or during, the respondents' participation in the survey, a process called priming \cite{Burton2019psychology}.  The possibility that the research participants are using gender bias when they respond to the questions cannot be eliminated. However, the only  gender-based cues in the survey are the gender of the participant. Thus, while the survey could prime participants for \emph{gender differences}, it is not priming participants for \emph{gender bias}.

The original purpose of collecting demographic data was to ensure that the survey respondents were a representative sample of the Australian cybersecurity industry as is good practice for  survey research \cite{Kel_et_al_2003}.  Preliminary analysis of the data \cite{HallRao2020} showed a disconnect between the non-technical skills sought by the cybersecurity industry, and the non-technical skills taught by Australian universities in their Master of cybersecurity programs.  Further examination of the data shows a distinct gender difference (based on the gender of the respondent/recruiter) in the non-technical skills  highly sought when hiring junior staff.

To delve deeper into the gender difference in the skills being sought, the non-technical skills were divided into \textit{people-focused} skills and \textit{task-focused} skills. 
Based on the socio-technical nature of cybersecurity \cite{schneier2012importance}, the following skills were classified as people-focused, i.e., skills requiring interactions with other people. Thompson \cite{thompson2021people} categorises many of these skills as \emph{interaction} skills, noting that working with people requires interaction. 

\begin{enumerate}
\item Inclusivity and Teamwork,
\item Verbal Communication, 
\item Written Communication,
\item Emotional Intelligence, 
\item Leadership.
\end{enumerate}

Leadership (classified as `working-with-others') is the only one of these skills explicitly appearing in the list of core skills in the most recent Future of Jobs report \cite{world2023future}.

The next set of  skills were classified as task-focused, i.e., skills that can be demonstrated in solitary situations, or in the context of other people.   Most of these skills are also referred to as \emph{personal effectiveness} \cite{thompson2021people}. 

\begin{enumerate}
\item Curiosity and Creativity, 
\item Confidence and Positive Attitude,  
\item Work Ethic and Personal Organisation,
\item Flexibility and Adaptability, 
\item Personal Integrity. 
\end{enumerate}

  \emph{Intervention} skills needed for problem-solving are listed in \cite{thompson2021people} but not considered  here.

\section{Results \label{sec:results}}
The results of the survey were collated and tabulated.  The research question was examined by disaggregating the responses to Questions A and B  using the demographic data.

\subsection{Survey of skills currently or recently used by cybersecurity practitioners}
First we examine the responses to Question A.

Question A: {\em In your current or most recent professional role,
what are the most important non-technical skills?}

\begin{figure}[ht!]
\includegraphics[width=.95\linewidth]{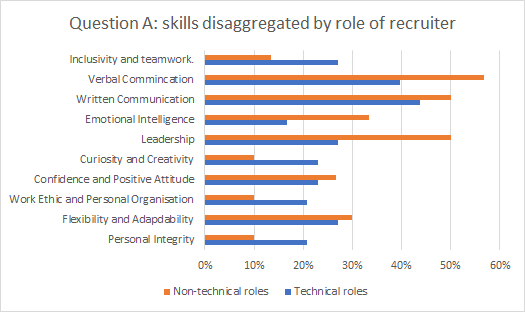}
\caption{The responses to Question A, disaggregated  by role \label{chartA_role}}
\end{figure}

\begin{figure}[ht!]
\includegraphics[width=.95\linewidth]{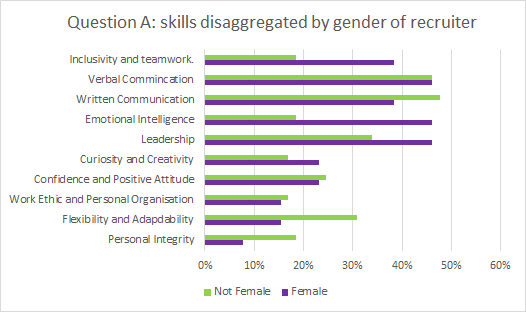}
\caption{The responses to Question A, disaggregated  by gender\label{chartA_gender}}
\end{figure}

The responses to Question A,  disaggregated by role, are shown in Figure \ref{chartA_role} while the same data, disaggregated by gender, is displayed in Figure \ref{chartA_gender}.  The skills surveyed are listed in the same order as in section \ref{subsec_dataanalysis}, with the people-focused skills in the top half of the charts and the task-focused skills in the lower half.

Neither of these charts show any distinct trends.   Statistical analysis was performed, and no significant trends were found in comparing the data by gender, role or skill grouping.

\subsection{Survey of skills sought when hiring junior cybersecurity staff}
Next the responses to Question B are examined.

Question B: {\em When recruiting junior staff into your team, what
non-technical skills are you most looking for?}

\begin{figure}[ht!]
\includegraphics[width=.95\linewidth]{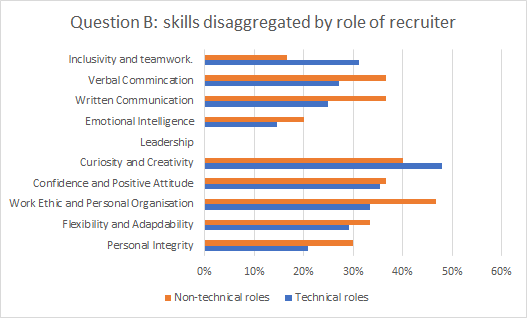}
\caption{Responses to Question B, disaggregated  by role \label{chartB_role}}
\end{figure}

\begin{figure}[ht!]
\includegraphics[width=.95\linewidth]{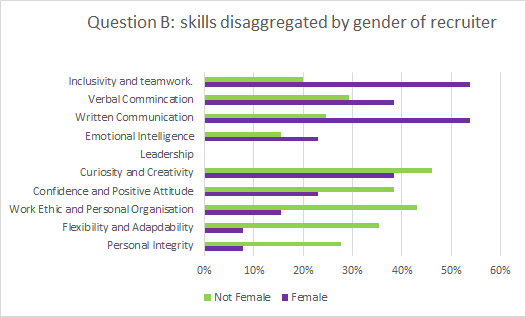}
\caption{Responses to Question B, disaggregated  by gender 
\label{chartB_gender}}
\end{figure}

The responses to Question B, disaggregated by role, are shown in Figure \ref{chartB_role} while the same data, disaggregated by gender, is displayed in Figure \ref{chartB_gender}.  Again, the skills surveyed are listed in the same order as in section \ref{subsec_dataanalysis}, with the people-focused skills in the top half of the charts and the task-focused skills in the lower half.

Figure \ref{chartB_role} shows no significant difference in the skills sought by technical vs non-technical professionals, whereas Figure \ref{chartB_gender} shows distinct differences in the skills sought by female vs non-female recruiters. Figure \ref{chartB_gender} indicates that female respondents value all people-focused skills more, whereas non-female respondents value  all  task-focused skills more.

The people-focused skill of `inclusivity and teamwork'  shows the strongest gender difference with over 50\% female respondents nominating this skill, whereas only 20\% of non-female respondents nominated it as important.

On the other hand, the task-focused skill of `curiosity and creativity' shows the least gender difference; 38\% for female and 46\% for non-female.

To further examine the gender differences  the nominations for each of the skills were tallied, and then split by gender, and by role.  There were 216 skill nominations, with some survey respondents only nominating one or two skills.

\vspace{2mm}
\begin{table}[ht!]
 \caption{Skills nominated by respondents, classified by gender and by role}
\begin{center}
 
\begin{tabular}{cccc}
& & People-Focused & Task-Focused\\
 \hline

 \multirow{2}{*}{Gender}
 & Female & 65\% & 35\%\\
  & Non-Female & 32\% & 68\%\\
 \hline
\multirow{2}{*}{Role}
& Technical & 37\% & 63\%\\
& Non-Technical & 37\% & 63\%\\\hline
 \end{tabular}
 \vspace{2mm}

\end{center}
\end{table}

No differences emerge when the data is examined by role, between the proportion of people-focused skills and task-focused skills nominated as important.

However, when examined by gender, there is a distinct difference. Female respondents nominated people-focused skills as important almost twice as often as task-focused skills, with the reverse being true for non-female respondents.  

The $p$ value for this difference, which is $p<0.005$, ($p = 0.0034$),
 demonstrates that even with this relatively small sample there is more than 99\% confidence that the difference is due to the gender of the respondents.  

The data indicates female cybersecurity practitioners highly value people-focused non-technical skills when recruiting into their teams, whereas non-females highly value task-focused non-technical skills. 

\section{Discussion }
The results of this study into the non-technical skills being sought when recruiting junior cybersecurity staff indicate that female cybersecurity practitioners  seek different skills to their non-female colleagues. The data indicates that when deciding the most important non-technical skills needed in a  recruit, the gender of the recruiter may be more important than the role of the recruiter. This fits within the established research of gender differences in leadership and management practices \cite{stelter2002gender,csahin2017leaders}.

\subsection{Role type v Gender}
Cybersecurity, as with many other technical industries, is heavily male dominated; only 25\% of the global cybersecurity workforce is female \cite{ISC2020} (17\% in Australia \cite{Risse2023}).  Non-technical roles within the cybersecurity industry, such as business development, HR, and sales, have more female staff, though these also are not yet gender balanced.   Some large technology companies have around 30\% female staff, with around 20\%  technical staff being female vs about 40\%  non-technical staff being female. \cite{Microsoft2020}.  The respondents to this survey represent a similar skewed distribution, with 13\% of those in technical roles being female vs 24\% of respondents in non-technical roles being female.

The gender differences in our data, with regards to skills sought when recruiting cybersecurity professionals into a team, cannot be explained by the different roles.  The data by role type does not indicate any statistically significant differences, meaning that the gender of the recruiter is the more important factor than the role type of the recruiter, with regards to skills being sought in a recruit. 

The lack of gender differences when a professional is reporting the most important skills needed to do their own role (Question A) is in stark contrast to that arising with regards to the skills they seek to add to their team (Question B). Can the impostor syndrome explain this difference? There is strong evidence that female professionals suffer from an imposter syndrome, often underrating their own skill level \cite{Chrousos749}.  The survey (Question A) asked  respondents  to report the skills they considered most important for their current (or more recent) role. Perhaps, a gender difference may have  appeared  if the survey had included a question asking  respondents to rate themselves on these important skills \cite{Strachan_etal_2018}. 

Hence, are the gender differences found in the data more related to the evaluation of personal skill level by respondents, and not in the understanding of the  most important skills? Are female respondents trying to fill a (perceived) gap, in themselves, or in their team? 

By classifying skills as either people-focused or task-focused, the research found strong evidence that female recruiters are more likely to look for staff with people-focused skills. A follow up study involving long form interviews in currently being conducted to examine the issues raised in further depth.

\subsection{Validation of study by comparison with multiple data sources}
 
  There are limitations to this study. Almost all large surveys conducted  (eg. \cite{ISC2022,world2023future}) draw conclusions based on self-reporting by professionals. The number of respondents to our survey is relatively small, and the data is all self-reported, limiting the statistical significance of the findings. Thus, the conclusions cannot be examined in isolation. 
  
  Hence, we examine the data  in the context of industry, government, and academic literature, to support our analysis.
 The skills used in the survey are sourced from both the Future of Jobs report \cite{world2018future} and from job advertisements. Job advertisements allow universities to keep in touch with graduate recruitment requirements, are often used at open days and information sessions for prospective students, and are  valid sources for industry skill requirements. 
 
 The World Economic Forum \cite{world2018future,world2023future} periodically asks companies globally about the skills these companies consider important for their current and future employees. The list of skills and their classifications does keep changing and are not specific to the cybersecurity industry. The skills in our survey came from the 2018  report \cite{world2018future} and form a good sample of what employers in general require.

 Professional bodies \cite{ISC2020, ISC2022} and governments \cite{AustCyber2018, ABS2021} collect and publish data on the demographics and characteristics of the Australian cybersecurity workforce.  These data sets are much more comprehensive, though collecting slightly different data than the small data set collected in this research.  There are also data sets on the characteristics of the  IT workforce \cite{Microsoft2020, NCWIT2020} or the STEM workforce \cite{burke2019labour, Industry2023}.  Based on these larger data sets, the survey respondents do indeed form a representative sample of the Australian cybersecurity industry.

Studies on gender bias in recruitment have mainly focused on the gender of the person being recruited \cite{moss2012science, Gorsuch2019gender, zhang2019gender}.  This study into the difference caused by the gender of the recruiter addresses a gap in the research on recruitment practices.
 Some recruitment firms suggest the use of diverse recruitment/selection panels \cite{Fuhl2020-eliminatinggenderbias}, but governments \cite{AusGov-Workplace2019} as yet do not. This research could help change that advice.

 \section{Conclusion}
There are many aspects of workplace diversity: gender, ethnicity, seniority, academic background, role, experience etc. 
Research shows that gender diversity on corporate decision-making teams leads to measurable differences in the profitability of the organisation. More importantly, diverse teams are better able to explore the non-standard ideas needed to tackle the challenges currently facing humanity. 

In this study, gender-based diversity was compared to role based diversity.  The analysis presented here indicates that in a recruitment situation, \textit{gender based diversity in the selection panel is more likely to provide alternative view points than role based diversity}.  

The inclusion of gender demographic data made this research possible. It found that females and non-females look for different skill sets when recruiting graduates in the cybersecurity field.  Females tend to focus on people skills while non-females tend to focus on task skills. With studies and research  showing that diverse teams make better decisions, having gender diversity in graduate recruitment teams should lead to better recruitment decisions. 

This study can be repeated, for other countries, and not just those with similar societal profiles to Australia. The questions asked focused on the non-technical skills used by recruiters themselves, and those sought in junior recruits. An extra question could be added asking respondents to self-select their (or their teams') perceived competency in the skills chosen.

A follow up study interviewing cybersecurity professionals is being conducted and should result in a more nuanced view of the gender differences highlighted in this paper.

\subsection*{Acknowledgment} Ethical Approval granted by RMIT College of Science Engineering and Health Human Research Ethics Advisory Network, project number 03-18/21934.

This research was supported by the RMIT Information Systems (Engineering) Enabling Capability Platform Grant CDF19068.

\end{document}